\documentclass[aps,prl, twocolumn, superscriptaddress]{revtex4-1}
\usepackage{graphicx}
\usepackage{amsfonts}
\usepackage{amsmath}
\usepackage{xcolor}

\begin{document}

\title{Confinement and collective escape of active particles }% from a trap}
\author{Igor S Aranson}
\affiliation{Departments of Biomedical Engineering, Chemistry, and Mathematics, Penn State University, University Park, Pennsylvania 16802, USA }
\author{Arkady Pikovsky}
\affiliation{Institute for Physics and Astronomy, University of Potsdam, Karl-Liebknecht-Strasse 24/25, 14476 Potsdam-Golm, Germany}
\affiliation{Department of Control Theory, Nizhny Novgorod State University, Gagarin Avenue 23, 606950 Nizhny Novgorod, Russia}

\date{\today}
\begin{abstract}
 Active matter broadly covers  the dynamics of self-propelled particles.    While the onset of collective behavior in homogenous active systems is relatively well  understood, the effect of inhomogeneities  such as obstacles and traps lacks overall clarity. Here, we study how interacting self-propelled particles become trapped and released from a trap. We have found that captured particles aggregate into an orbiting  condensate with a crystalline structure. As more particles are added, the trapped condensate escape as a whole. Our results shed light on the effects of confinement and quenched disorder in active matter. 
    \end{abstract}
\maketitle
%\section{Introduction}
Assemblies of interacting self-propelled particles, broadly defined as active matter, continue attracting significant attention \cite{bechinger2016active,aranson2013active,gompper20202020}. 
In the last ten years, notable progress was achieved in the understanding of the onset of collective behavior \cite{chate2020dry,peshkov2012nonlinear,patelli2019understanding} and characterization of some collective states \cite{solon2015pressure,alert2020universal}. Spatial inhomogeneities,  surface roughness, or  quenched disorder  play a significant role in active systems \cite{peruani2018cold,duan2021breakdown,ro2021disorder,olsen2021active}. Quenched disorder, for example, may lead to the onset of trapped states, anomalous diffusion,  and breakdown of ergodicity. 

The motion of self-propelled particles on a disordered substrate is relevant in the context of the ``active conductivity''. This situation is realized, for example, when motile bacteria propagate through a porous environment \cite{waisbord2021fluidic}. 
It is an analog of the equilibrium problem of electrons migration in random media. If the substrate is approximated as an array of well-separated traps (impurities), then  the process can be viewed as a sequence of  escapes and re-trappings of particles. In the context of electrons trapped by impurities, a non-zero conductance is due to the overlap of wave-functions of electrons at individual impurities (Dykhne theorem \cite{bychkov1971theory}). However, no such general result is known for active particles. 

In this Letter, we study how self-propelled particles are captured  and released by an isolated trap modeled as a potential well. We start with non-interacting particles and show that an 
individual particle typically exhibits  chaotic scattering by a trap. Then we introduce interactions.  A Lennard-Jones potential 
organizes particles  into an orbiting condensate with the hexatic crystalline order. 
An alignment coupling brings dissipation and synchronization
in the dynamics,
and results in a perpetual capture of active particles: a trap 
becomes an analog of a ``black hole''. 
``Bombardment'' by active particles results in particle absorption, condensate melting,  and  recrystallization. Above a certain threshold number of captured particles, the trap storing capacity is exceeded, and the condensate escapes as a whole. 

Active particles in a harmonic trap have been studied 
both experimentally \cite{dauchot2019dynamics,schmidt2021non,takatori2016acoustic} and theoretically  \cite{pototsky2012active,hennes2014self,wexler2020dynamics,das2018confined,jahanshahi2017brownian}. The main focus was on the steady-state distributions or  escape of individual particles due to the combined effect of self-propulsion and thermal fluctuations. In this Letter, we consider purely deterministic effects of propulsion and interactions;
this aspect of our model manifests a crucial difference not explored in other publications.   

%\section{An active particle in a trapping potential}
%\subsection{General force}
We consider a self-propelled particle  moving in two dimensions with a constant velocity
$V$,  while a force $\mathbf{f}$ acting on the particle only rotates its
direction of motion:
\begin{equation}
\frac{d{\bf r}}{dt}=\tilde V {\bf n}\;,\qquad \frac{d{\bf n}}{dt}=\mathbf{f}-({\bf n}\cdot\mathbf{f}){\bf n}\;.
\label{eq:genforce}
\end{equation}
Here ${\bf n}=(\cos\theta,\sin\theta)$ is the unit vector in the direction of motion. The equation for $\bf n$ ensures that $|{\bf n}|^2=1$. Thus, only the velocity direction changes and obeys $\dot\theta=f_y\cos\theta-f_x\sin\theta$.
Physically, velocity aligns with the potential gradient. This situation can be realized, for example, for magnetic particles in a magnetic trap. Models of this type were also discussed in the context of artificial chemotaxis  \cite{liebchen2016pattern,liebchen2018synthetic,stark2018artificial}. 

%\subsection{Two-dimensional potential}
Below
we consider different types of forces, but we start with a motion
in an external potential field ${U}(x,y)$, so that $\mathbf{f}_{\rm pot}=-\nabla U$.
If we renormalize variables so that the width and the depth of the potential are one, the only remaining relevant parameter is the dimensionless velocity $V$. The limit of strong potential is that of $V\to 0$. 
%A weak potential corresponds to large values of velocity $V$. 
The corresponding equations can be written as 
\begin{eqnarray}
%\begin{gather}
\frac{d {\bf  r } }{dt }&=& V {\bf n}\;, \label{eq:beqxy} \\
\frac{d \theta}{dt } &=& - \partial_y U\cos\theta+\partial_x U\sin\theta\;. \label{eq:beqth}
%\end{gather}
%\label{eq:beq}
\end{eqnarray}
%\textcolor{red}{AP: I changed here because I think writing the equation for $\theta$ via vector product is somehow cumbersome $\dot\theta=\mathbf{e}_z\cdot (\mathbf{n}\times (-\nabla U))$}

%\subsection{Hamiltonian}
Equations  (\ref{eq:beqxy}), (\ref{eq:beqth})  possess 
 the Hamiltonian 
\begin{equation}
H({\bf r},{\bf p})=V |{\bf p}|-\exp\left[-\frac{U({\bf r})}{V}\right]=0\;.
\label{eq:ham}
\end{equation}
Because of the energy conservation,  the relation $|{\bf p }|=V^{-1}\exp[-U({\bf r}) V^{-1}]$ holds. A substitution 
${\bf p} = V^{-1} \exp[-U({\bf r}) V^{-1}]{\bf n}$ reduces the Hamiltonian dynamics to Eqs.  (\ref{eq:beqxy},\ref{eq:beqth}). 
%$p_x=V^{-1}\exp[-U(x,y) V^{-1}]\cos\theta$,
%$p_y=V^{-1}\exp[-U(x,y) V^{-1}]\sin\theta$. 
%Then the  Hamiltonian equations according to \eqref{eq:ham} reduce to Eqs.  (\ref{eq:beqxy}), (\ref{eq:beqth}). 
%Eqs. (\ref{eq:beqxy}), (\ref{eq:beqth}) 
The Hamiltonian  \eqref{eq:ham} coincides with  that describing ray propagation in geometrical optics, with the refraction index $\sim \exp[-U({\bf r}) V^{-1}]$ \cite{Kravtsov-Orlov-90}.
%\subsection{Trapping in a potential well}

In
the  small velocity limit,  $V \ll 1 $, one can separate 
slow migration of the particle (Eq.~\eqref{eq:beqxy}) and its fast alignment with the gradient of the potential (Eq.~\eqref{eq:beqth}).  The orientation angle $\theta$ in Eq. 
\eqref{eq:beqth} adjusts to the gradient angle 
 $\alpha = \arctan(\partial_y U/\partial_x U)$ as
$
\dot\theta=|\nabla U|\sin(\alpha-\theta)
$.
This fast adjustment of the direction of motion  toward the minimum of the potential is followed by a slow drift~\eqref{eq:beqxy}.  Once the vicinity of the potential minimum is reached, 
 i.e.,
$|\nabla U|\approx 0$, the scale separation breaks down, and one has to  consider
the full Eqs.~\eqref{eq:beqxy},\eqref{eq:beqth}.

%To see what happens near the minimum of the potential, it
%is enough to consider the motion in
%a quadratic potential, which 
%is generically asymmetric:
In the vicinity of a  minimum, a generic potential is approximated by an asymmetric parabolic well, 
$U_{\rm pb}=(x^2+b^2 y^2)/2$. Since the
depth of this harmonic potential is not defined,  one can set
by virtue of a renormalization,
parameter $V$ in Eqs.~\eqref{eq:beqxy},\eqref{eq:beqth} to one. Then the only parameter left  
is the potential asymmetry  $b$. Since the  dynamics is Hamiltonian, the type of motion depends on
initial conditions.

 \begin{figure}
 \centering
 \includegraphics[width=\columnwidth]{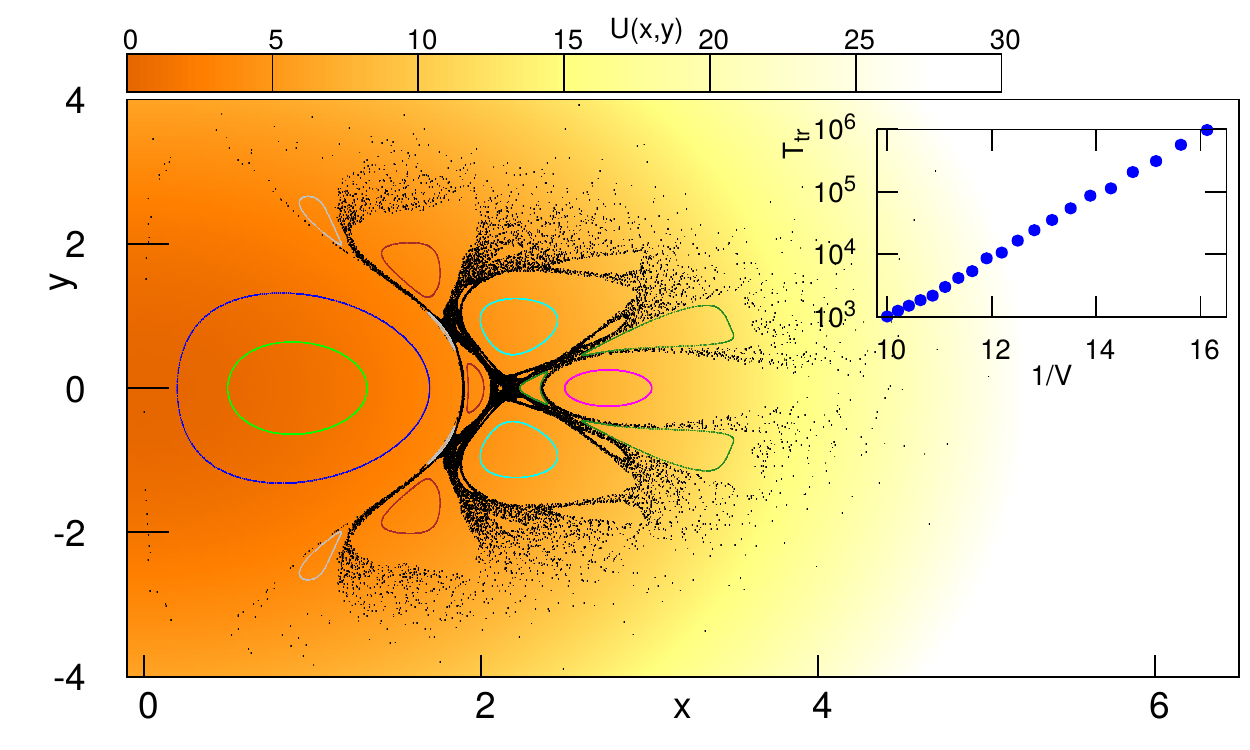}
 %(b)\includegraphics[width=0.8\columnwidth]{gauss_stat_1.pdf}
\caption{ 
Poincar\'e map  for a parabolic potential  $b^2=\sqrt{5}-1$ (color background).  Black dots: a chaotic trajectory starting at large $|x|,|y|$. 
The  chaotic dynamics is not ergodic as it  contains voids with  closed curves - images of quasiperiodic trapped motion. 
%Some of these sets have several branches.
Inset: Trapping time   $T_{\rm tr}\sim \exp(1/V)$ in a Gaussian potential well  vs velocity $V$, $b=(\sqrt{5}-1)/2$.  
%(c) Decay time for a chaotic molecule of build of $N$ active particles with %Lennard-Jones interation.
}
\label{fig:parp}
\end{figure}

Figure~\ref{fig:parp} illustrates the dynamics 
for  $b^2=\sqrt{5}-1$ (other values of $b$ give qualitatively the same
picture). The three-dimensional phase space is
reduced to  a two-dimensional Poncar\'e map (we chose  the section
$\cos\theta=0$, $\frac{d}{dt}(\cos\theta) <0$). One  sees a large
chaotic domain (black dots)~(video 1 in \cite{suppl}) and regular islands filled with quasiperiodic trajectories. 
%The points of the chaotic trajectory 
%are concentrated at small values
%of $x,y$, where these coordinates are of the order of one. 
%In
%the original representation, where $V$ is a parameter, this corresponds
%to a domain of size $V^{1/2}$. 
This Poincar\'e map is unusual for the Hamiltonian dynamics: the distribution is very inhomogeneous -- density of points at small $\mathbf{r}$ values  is much larger than at large ones.  
%of these variables.
That happens because the ``natural'' coordinates $({\bf r},\theta)$
are not the canonical ones. One  estimates the density
on the plane $(x,y)$ by assuming a fully developed chaos where the angle $\theta$ is random and uniformly distributed. Then integration of
a microcanonical distribution density for the Hamiltonian \eqref{eq:ham} $w({\bf r},{\bf p})\sim \delta\left(V |{\bf p}| -\exp\left[-\frac{U({\bf r})}{V}\right]\right)$  over the angle $\theta$ yields 
$W({\bf r} ) \sim\exp\left[-U({\bf r})/{V}\right]$.
This  resembles the Gibbs-Boltzmann distribution, with the velocity
$V$ playing a role of the temperature. This result implies that
the chaotic motion in Fig.~\ref{fig:parp} spreads to arbitrarily
large values of the potential, although it is rather improbable to reach these heights. We conclude that
although  a slow particle arrives
at the minimum of the potential and moves chaotically there, 
after a  long  time, it returns to the high values of the 
potential where it started.  Such a return must happen  according to the recurrence  
of the Hamiltonian trajectories. In addition to a chaotic region in Fig.~\ref{fig:parp}, there are domains
of quasi-periodic  dynamics concentrated 
close to the potential minimum. This  motion can occur  for particles starting close to the minimum. % but not for those coming from outside regions.    

%The properties of the particle motion outlined above imply
%that in 

The Hamiltonian structure of Eqs.  \eqref{eq:beqxy},\eqref{eq:beqth}
implies that capture of particles falling in
a finite-depth potential well  is impossible.  Only temporary trapping occurs that can be interpreted as a chaotic scattering: a particle falls into the well, goes to its minimum, and moves there chaotically like in Fig.~\ref{fig:parp}, but eventually rises high and escapes (video 2 in \cite{suppl}). 
Furthermore, because of the exponential density $W\sim\exp[-U/V]$,  the characteristic trapping time obeys a Kramers-like law $T_{\rm tr}\sim \exp[-1/V]$. This is confirmed in Fig.~\ref{fig:parp}(inset)
where  the mean trapping time on a Gaussian  potential well
$U(x,y)=-\exp(-x^2-b^2 y^2)$ is shown. 
%This potential has unit depth and width of the order one, according to our normalization.

%\section{Interacting particles}
We consider two types of interactions between particles below: (i)  interaction  by a potential force; (ii) an alignment to an average (over a neighborhood) orientation of neighboring particles; 
this latter interaction is of the Vicsek (or Kuramoto) 
type \cite{Vicsek_etal-95,chate2020dry}. 
%\subsection{Conservative coupling: Lennard-Jones potential}

 \begin{figure}
 \centering
 \includegraphics[width=\columnwidth]{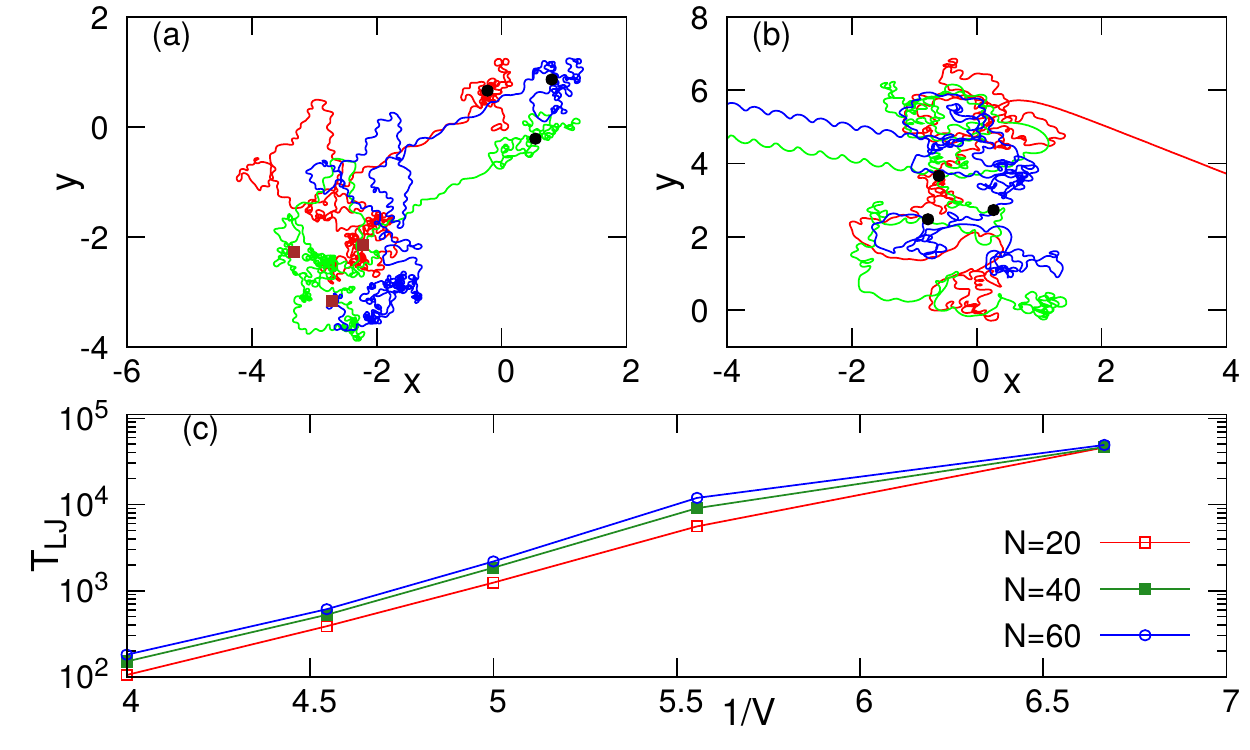}
\caption{Trajectories of three coupled particles
in $(x,y)$ plane, for the LJ potential, for $V=0.1$ (a)
and $V=0.2$ (b). Circles show positions of 
particles at time $t=100$, squares at time $t=500$. In panel (b) 
one  sees the breakdown of   the bounded state  into a couple of particles moving to the left, and a particle moving to the right. (c) Average lifetime of a crystal of $N$  particles 
vs inverse velocity $1/V$. The dissociation time is defined as an event when one particle leaves. 
}
\label{fig:lj3}
\end{figure}

We start with the interaction via a conservative pair-wise  force that
depends on the distance between the particles.  This case, like the 
motion in the external potential above, is Hamiltonian. 
We explore below the Lennard-Jones (LJ) potential 
$U_{\rm LJ}({\bf r}_1,{\bf r}_2)=4(\rho^{-12}-\rho^{-6})$, where 
$\rho^2=|{\bf r}_1-{\bf r}_2|^2$. Several particles having a small velocity $V$ placed close to each other, 
form a bounded state due to the LJ-coupling. 
%Its features depend on the number of particles and on the velocity parameter $V$.
For two particles, the bounded state is quasiperiodic, while for $N\geq 3$ 
it is typically chaotic. Several particles form a chaotically vibrating
``crystalline molecule'' with a hexatic order. The center of mass (c.o.m.) performs a diffusive motion in the plane $(x,y)$, and the particles from time to time rearrange their positions, see Fig.~\ref{fig:lj3}(a,b) and Video 4 in \cite{suppl}. Because the  LJ potential is finite,
there is a non-zero probability for a particle to escape. 
Such an event is shown in Fig.~\ref{fig:lj3}(b). 
Figure~\ref{fig:lj3}(c) shows that the lifetime $T_{\rm LJ}$ of 
a ``shaking crystal'' without confining potential 
has the same Kramers-type dependence on the 
velocity $T_{\rm LJ}\propto \exp[V^{-1}]$, as the 
lifetime of a particle in a potential well (cf. Fig.~\ref{fig:parp}(inset)).
A crystalline molecule has a significant lifetime only for particles with $V\lesssim 0.5$.

%\subsection{Alignment coupling}
The aligning force $\mathbf{F}_k$ acts in the direction 
of the weighted average of the velocities of other particles in a  neighborhood  of $k$-th particle:
\begin{equation}
    \mathbf{F}_k=\epsilon\sum_j S(|\mathbf{r}_j-\mathbf{r}_k|) \mathbf{n}_j\;.
\label{eq:alforce}    
\end{equation}
The distance-dependent factor (we assume it to be Gaussian $S(r)=\exp(-r^2/r_0^2)$) defines the range $r_0$
of the force.  
%We assume a Gaussian form $S(r)=\exp(-r^2/r_0^2)$.
%Here $r_0$ is the characteristic size of the alignment neighborhood. 
Parameter $\epsilon$ determines the strengths of the alignment. The alignment force is velocity-dependent and dissipative. 
In terms of the velocity direction $\theta$ it has the form of Kuramoto-type coupling $\dot\theta_k\sim \sin(\theta_j-\theta_k)$. 
%Therefore,   it has the form of Kuramoto-type coupling (if  $\theta$ is interpreted as a phase).

We examine next  a combination of the alignment and 
the conservative forces due to a confining  potential or an LJ interaction
(for  a pure alignment
 see~\cite{Kruk_etal-18}). 

\begin{figure}
 \centering
 \includegraphics[width=\columnwidth]{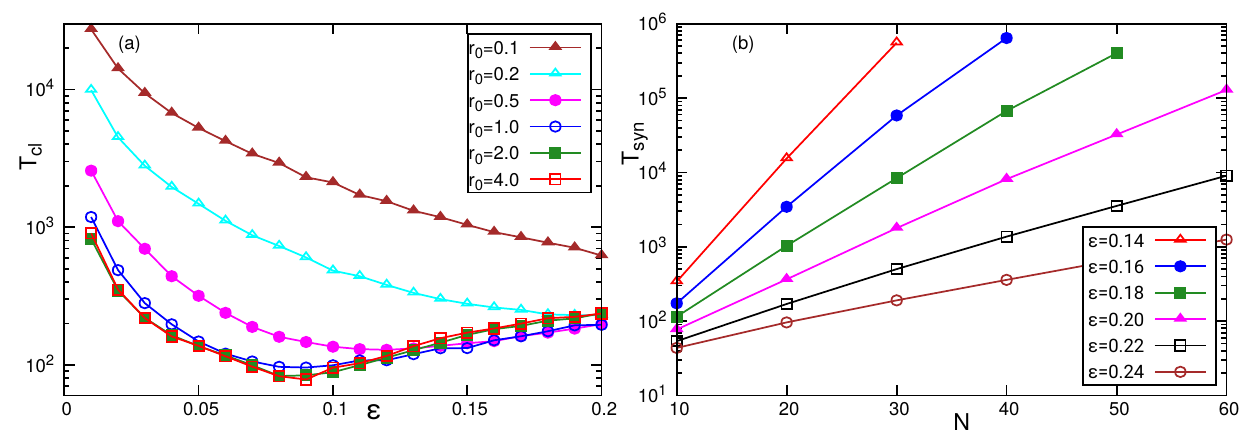}
\caption{ (a) The average time at which 20 initially randomly distributed particles
in a harmonic potential form 10 clusters due to alignment. For small $r_0$ this
time monotonously decreases with coupling strength $\epsilon$, while large ranges of  $r_0\gtrsim 0.5$
there is an optimal coupling strength. (b)
Average time to synchronization for chaotic crystals composed of $N$ particles,
for $V=0.1$ and different strengths of alignment forces $\epsilon$.
}
\label{fig:hk}
\end{figure}

We start with a set of chaotic particles in a harmonic trap  
(Fig.~\ref{fig:parp}), described by 
Eqs.~\eqref{eq:beqxy},\eqref{eq:beqth} with additional 
alignment \eqref{eq:alforce}. 
The main observation is that 
for large values
of $\epsilon$ and large ranges of coupling $r_0$, 
particles always synchronize:
 after some transient time,
all the coordinates and angles coincide, and the particles
form a synchronous point cluster ${\bf r}_1=\ldots={\bf r}_N$, 
$\theta_1=\ldots=\theta_N$ (video 3 in \cite{suppl}). This synchronization
is possible because the aligning force is 
dissipative \cite{peruani2018cold}.
In the final synchronous state dissipation disappears
(the  force $\sim\sin(\theta_j-\theta_k)$ vanishes),
and the trajectory of the cluster is described again by
the Hamiltonian dynamics \eqref{eq:beqxy},\eqref{eq:beqth}. However, in the course 
of  alignment, particles leave the chaotic domain, and the final dynamics is 
quasiperiodic (cf. Fig.~\ref{fig:parp}). Thus, strong
alignment  synchronizes particles and regularizes their motion. For 
lower alignment  rates $\epsilon$ and especially for small ranges $r_0$, 
multiple states with several clusters are observed up to large 
times. If the alignment coupling radius $r_0$ is small enough, several regular 
clusters
may effectively stop to interact; then they constitute an ``attractor''.  Noteworthy, standard methods of the synchronization theory, like the master stability function method \cite{pecora1998master}, are not applicable here because the type of motion (from chaos to quasiperiodicity) changes over time. In Fig.~\ref{fig:hk}(a) we illustrate the rate of synchronization in dependence on the parameters $\epsilon$ and $r_0$. 
At large
$r_0$ there is an ``optimal'' coupling strength. We attribute this to the fact that, for large $r_0$, the alignment coupling is a global one.  Because 
the velocities of distant particles are effectively ``de-aligned'' by different
potential forces they experience at different positions in the harmonic trap, the alignment slows down. At small values of $r_0$, only neighboring particles interact, and they are much more ``synchronizable'' because their trajectories in the potential may easily adjust as well.

In the case of alignment of particles trapped by a finite-depth potential well,  two time scales compete:  the trapping time $T_{\rm tr}$  determined by the velocity $V$ (cf. Fig.~\ref{fig:parp}(inset));  the synchronization time
$T_{\rm s}$ (e.g., one can take the clustering time of Fig.~\ref{fig:hk}(a)). If
$T_{\rm s}\gg T_{\rm tr}$, almost all the particles escape 
from the well and spread. In the opposite limit, 
$T_{\rm s}\ll T_{\rm e}$,  a \textit{complete self-trapping} due to
alignment occurs:  a cluster of synchronous trapped quasiperiodically moving particles  appears. 
Thus, the potential well becomes an effective ``black hole'': particles are 
trapped perpetually due to dissipative alignment. 
In an intermediate case $T_{\rm s} \sim T_{\rm tr}$, some particles escape 
while others
form a perpetually trapped cluster.

%\subsection{Alignment of LJ-interacting particles}

A combination of the LJ and the alignment interactions  allows for synchronization of the crystal. Point clusters cannot be formed because particles 
cannot come close to each other due to the LJ repulsion at small distances. Thus,  only  orientations  $\theta_k$ can 
synchronize. Here again, a relation between the synchronization time and the
lifetime due to potential forces is crucial.

We describe the combined action of the LJ and alignment interactions for  particles in the finite-well potential, because,  for small enough
velocities, the particles are practically trapped forever. We chose the width of the well to be approximately ten times larger than the characteristic spatial scale of the LJ potential so that only a few particles fit into the well.

 \begin{figure}
 \centering
\includegraphics[width=\columnwidth]{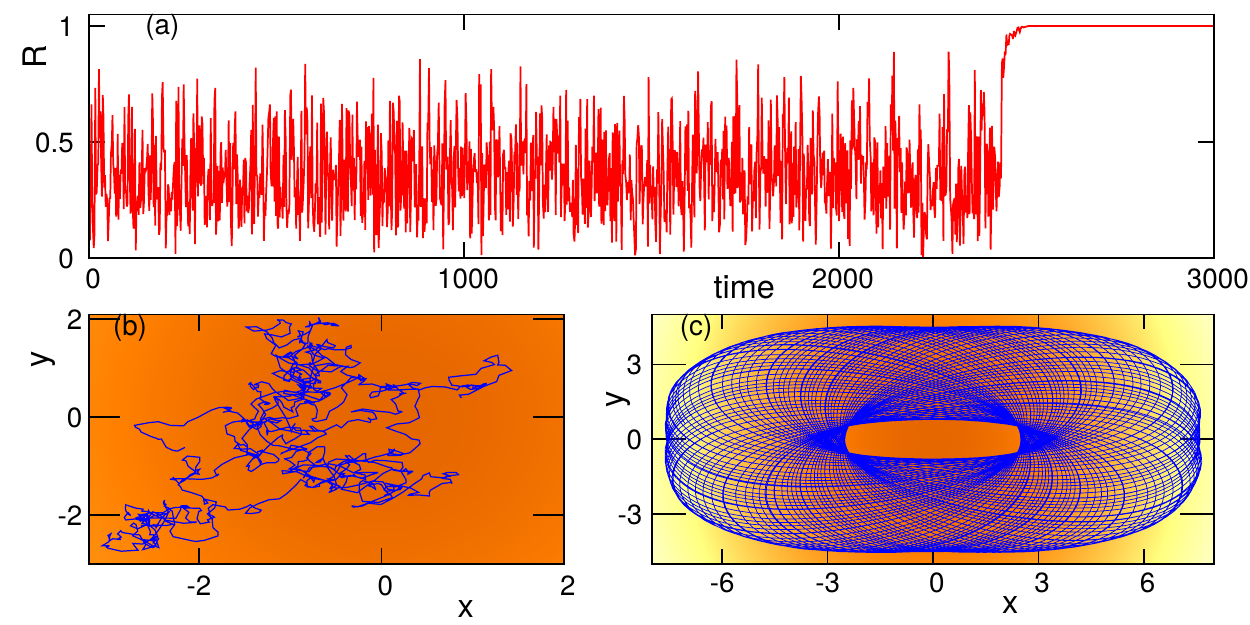}
\caption{Seven particles with a LJ potential and alignment in a potential well (color coding). Panel (a): the order parameter $R$ vs time.
Panels (b,c): trajectories of the c.o.m  on the $(x,y)$ plane in the disordered  $t<2300$ (b) and in the ordered  $t>3000$ (c) domains.
}
\label{fig:gklj}
\end{figure}

Figure~\ref{fig:gklj}  illustrates the dynamics of seven particles.
Initially, they form a shaking crystal with a random motion (reminiscent of a confined random walk) of the c.o.m, Fig.~\ref{fig:gklj}(b).  The orientational order is characterized by the Kuramoto order parameter $R=|\langle e^{i\theta}\rangle|$, Fig.~\ref{fig:gklj}(a).
In the disordered state, the order parameter 
fluctuates around $R\approx 0.4$. At $t\approx 2500$, an 
abrupt transition to synchrony in the directions of motion is observed, 
beyond this transition $R\approx 1$. The crystal becomes ordered, 
and the c.o.m. performs
a quasiperiodic motion in the well, Fig. ~\ref{fig:gklj}(c), and  all the particles become perpetually trapped (video 5 in \cite{suppl}). 

The abrupt transition to synchrony from a chaotic crystal should be contrasted to the clustering transition without the LJ coupling. 
In the latter case, the order parameter grows gradually as the particles
continuously come closer and merge.  In contradistinction,  the process depicted in  Fig. ~\ref{fig:gklj}(a) is characterized as transient chaos that abruptly ends in an absorbing synchronized state (see a general exposition of transient chaos~\cite{Lai-Tel-11}, and a case of chiral active particles in  \cite{Pikovsky-21a}). 

To examine the dependence of the synchronization time
on the size of the crystal, we consider a set of $N$ active particles.
We take the  LJ interaction and additional alignment coupling with $r_0=1$ (i.e.,
the same range as the LJ potential)
 but without a confining external potential. To ensure that the lifetime of the crystal
is larger than a characteristic synchronization time (cf. Fig.~\ref{fig:lj3}(c)),
we considered particles with small velocity $V=0.1$.
Figure \ref{fig:hk}(b) shows that the dependence of the synchronization time
on the size of the crystal
is exponential $T_{\rm syn}\sim g(\epsilon)\exp[h(\epsilon) N]$, with
$\epsilon$-dependent factors $g,h$. That implies that here a super 
transient behavior~\cite{Lai-Tel-11, Pikovsky-21a} is observed, for which a characteristic 
time exponentially grows with the system size.

 %\begin{figure}
 %\centering
%\includegraphics[width=\columnwidth]{eval_slr_3.pdf}
%\caption{Average time to synchronization for chaotic crystals composed of $N$ %particles,
%for $V=0.1$ and different strengths of alignment forces $\epsilon$. 
%}
%\label{fig:slr}
%\end{figure}

 \begin{figure}[!tbh]
 \centering
\includegraphics[width=\columnwidth]{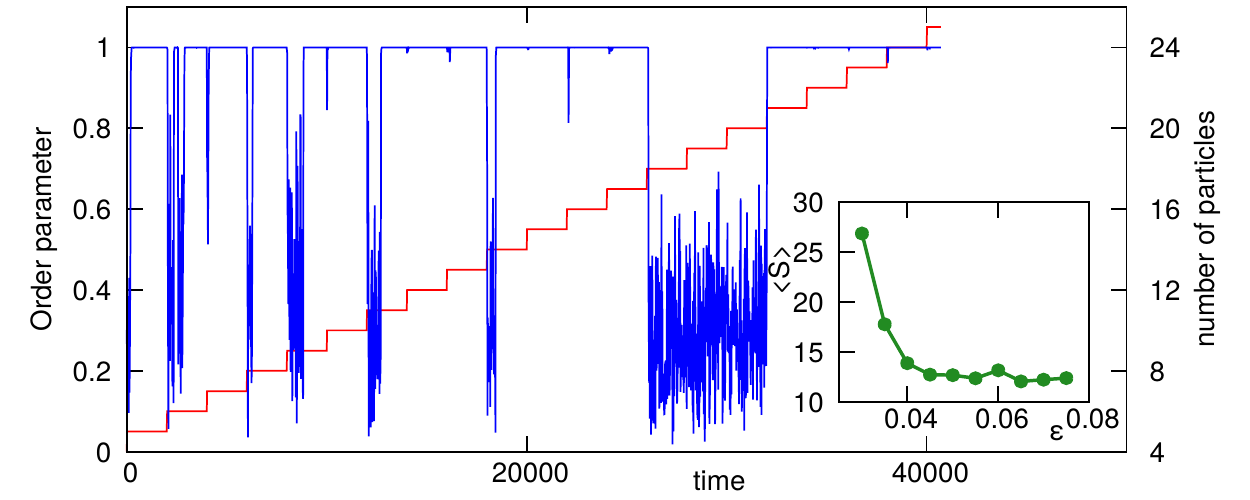}
 %(b)\includegraphics[width=0.9\columnwidth]{eval_escape_1.pdf}
 \caption{An example of trapping dynamics for $\epsilon=0.04$ and $V=0.15$ (a particle is added every 2000 time units). Blue line: evolution of the order parameter (left axis) showing domains of disorder and synchrony ($R\approx 1$). Red curve: number of particles (right axis). Inset: the average size of the escaped cluster vs  $\epsilon$. The time interval is proportional to
the size of the cluster because a particle is added every 2000 time units.
}
\label{fig:esc2}
\end{figure}

Finally, we studied  how many particles can be trapped
in a potential well
in the process illustrated in Fig.~\ref{fig:gklj}. We assume for simplicity that $r_0=\infty$, i.e., all particles in the well equally contribute to the alignment. 
The particles from outside ``bombard'' the trap  one-by-one at regular time intervals. They
form a crystal, the dynamics of which
is followed using the alignment order parameter $R$, see 
%A typical time dependence is illustrated in 
Fig.~\ref{fig:esc2}.
One  sees that in some cases adding  a particle melts the crystal. In other cases,  a particle is absorbed into 
the existing order. Finally, the crystal as a whole
escapes the potential well because  an incoming particle  ``kicks''
it from a trapped quasiperiodic regime into an escaping trajectory. 
We observed that only a synchronized crystal can escape (video 6 in \cite{suppl}). 
Fig.~\ref{fig:esc2}(inset) depicts the average size of the escaping
crystal $\langle S\rangle$ vs $\epsilon$. It shows a significant increase of a typical escaping crystal size for low alignment rates $\epsilon$. That occurs because for global coupling,  
the alignment rate is proportional to the number of particles. Therefore,   for small
$\epsilon$,  a sufficient number of interacting particles is required to  form a synchronous crystal.

%\section{Conclusion}
In conclusion, we investigated the trapping of individual and interacting active particles. The problem is nontrivial for slow particles (deep potentials), that spend a long time close to the bottom. We have found that non-interacting particles can be trapped
only for a finite  time due to the Hamiltonian structure of the 
equations of motion. In turn, an alignment of 
particles brings dissipation and establishes a time arrow. So, 
the  potential well becomes a ``black hole'' with perpetually captured particles on quasiperiodic trajectories. 
The particles form either a cluster or a coherent crystal if 
additional LJ coupling is taken into account. With the particle 
bombardment,
we observe a highly nontrivial trapping behavior: 
melting and re-synchronization of crystals, and eventually a coherent escape of the entire assembly. 

The problem we have studied is relevant for the understanding of active matter subject to quenched environmental disorder. For the media, which can be interpreted as an array of randomly positioned traps, the process
can be represented as a sequence of the described above trapping events.  Our study also shows that the disorder may have a finite trapping capacity. Once the traps are filled up, the further bombardment  may lead to the spontaneous avalanche-like release of many particles.
While this phenomenon is reminiscent of  self-organized criticality (SOC) \cite{bak1988self}, there are many fundamental differences:  the trapped states are highly dynamic and form chaotically or coherently moving crystals. Likely, this will result in a  different distribution for the avalanche sizes and other statistical characteristics. 
%Also, the dynamic nature of the processes results in  memory effects and dependence on the initial conditions.

\begin{acknowledgments}
We thank S. Klapp and H. Stark for useful discussions. A. P. was supported by the Russian Science Foundation, grant  17-12-01534, and by German Science Foundation, grant   PI 220/22-1.
Research of I.S.A. was supported by the U.S. Department of Energy, Office of Science, Basic Energy Sciences, under Award DE-SC0020964.
\end{acknowledgments}
%\bibliography{ap.bib}

\begin{thebibliography}{33}%
\makeatletter
\providecommand \@ifxundefined [1]{%
 \@ifx{#1\undefined}
}%
\providecommand \@ifnum [1]{%
 \ifnum #1\expandafter \@firstoftwo
 \else \expandafter \@secondoftwo
 \fi
}%
\providecommand \@ifx [1]{%
 \ifx #1\expandafter \@firstoftwo
 \else \expandafter \@secondoftwo
 \fi
}%
\providecommand \natexlab [1]{#1}%
\providecommand \enquote  [1]{``#1''}%
\providecommand \bibnamefont  [1]{#1}%
\providecommand \bibfnamefont [1]{#1}%
\providecommand \citenamefont [1]{#1}%
\providecommand \href@noop [0]{\@secondoftwo}%
\providecommand \href [0]{\begingroup \@sanitize@url \@href}%
\providecommand \@href[1]{\@@startlink{#1}\@@href}%
\providecommand \@@href[1]{\endgroup#1\@@endlink}%
\providecommand \@sanitize@url [0]{\catcode `\\12\catcode `\$12\catcode
  `\&12\catcode `\#12\catcode `\^12\catcode `\_12\catcode `\%12\relax}%
\providecommand \@@startlink[1]{}%
\providecommand \@@endlink[0]{}%
\providecommand \url  [0]{\begingroup\@sanitize@url \@url }%
\providecommand \@url [1]{\endgroup\@href {#1}{\urlprefix }}%
\providecommand \urlprefix  [0]{URL }%
\providecommand \Eprint [0]{\href }%
\providecommand \doibase [0]{http://dx.doi.org/}%
\providecommand \selectlanguage [0]{\@gobble}%
\providecommand \bibinfo  [0]{\@secondoftwo}%
\providecommand \bibfield  [0]{\@secondoftwo}%
\providecommand \translation [1]{[#1]}%
\providecommand \BibitemOpen [0]{}%
\providecommand \bibitemStop [0]{}%
\providecommand \bibitemNoStop [0]{.\EOS\space}%
\providecommand \EOS [0]{\spacefactor3000\relax}%
\providecommand \BibitemShut  [1]{\csname bibitem#1\endcsname}%
\let\auto@bib@innerbib\@empty
%</preamble>
\bibitem [{\citenamefont {Bechinger}\ \emph {et~al.}(2016)\citenamefont
  {Bechinger}, \citenamefont {Di~Leonardo}, \citenamefont {L{\"o}wen},
  \citenamefont {Reichhardt}, \citenamefont {Volpe},\ and\ \citenamefont
  {Volpe}}]{bechinger2016active}%
  \BibitemOpen
  \bibfield  {author} {\bibinfo {author} {\bibfnamefont {C.}~\bibnamefont
  {Bechinger}}, \bibinfo {author} {\bibfnamefont {R.}~\bibnamefont
  {Di~Leonardo}}, \bibinfo {author} {\bibfnamefont {H.}~\bibnamefont
  {L{\"o}wen}}, \bibinfo {author} {\bibfnamefont {C.}~\bibnamefont
  {Reichhardt}}, \bibinfo {author} {\bibfnamefont {G.}~\bibnamefont {Volpe}}, \
  and\ \bibinfo {author} {\bibfnamefont {G.}~\bibnamefont {Volpe}},\
  }\href@noop {} {\bibfield  {journal} {\bibinfo  {journal} {Reviews of Modern
  Physics}\ }\textbf {\bibinfo {volume} {88}},\ \bibinfo {pages} {045006}
  (\bibinfo {year} {2016})}\BibitemShut {NoStop}%
\bibitem [{\citenamefont {Aranson}(2013)}]{aranson2013active}%
  \BibitemOpen
  \bibfield  {author} {\bibinfo {author} {\bibfnamefont {I.~S.}\ \bibnamefont
  {Aranson}},\ }\href@noop {} {\bibfield  {journal} {\bibinfo  {journal}
  {Physics-Uspekhi}\ }\textbf {\bibinfo {volume} {56}},\ \bibinfo {pages} {79}
  (\bibinfo {year} {2013})}\BibitemShut {NoStop}%
\bibitem [{\citenamefont {Gompper}\ \emph {et~al.}(2020)\citenamefont
  {Gompper}, \citenamefont {Winkler}, \citenamefont {Speck}, \citenamefont
  {Solon}, \citenamefont {Nardini}, \citenamefont {Peruani}, \citenamefont
  {L{\"o}wen}, \citenamefont {Golestanian}, \citenamefont {Kaupp},
  \citenamefont {Alvarez} \emph {et~al.}}]{gompper20202020}%
  \BibitemOpen
  \bibfield  {author} {\bibinfo {author} {\bibfnamefont {G.}~\bibnamefont
  {Gompper}}, \bibinfo {author} {\bibfnamefont {R.~G.}\ \bibnamefont
  {Winkler}}, \bibinfo {author} {\bibfnamefont {T.}~\bibnamefont {Speck}},
  \bibinfo {author} {\bibfnamefont {A.}~\bibnamefont {Solon}}, \bibinfo
  {author} {\bibfnamefont {C.}~\bibnamefont {Nardini}}, \bibinfo {author}
  {\bibfnamefont {F.}~\bibnamefont {Peruani}}, \bibinfo {author} {\bibfnamefont
  {H.}~\bibnamefont {L{\"o}wen}}, \bibinfo {author} {\bibfnamefont
  {R.}~\bibnamefont {Golestanian}}, \bibinfo {author} {\bibfnamefont {U.~B.}\
  \bibnamefont {Kaupp}}, \bibinfo {author} {\bibfnamefont {L.}~\bibnamefont
  {Alvarez}},  \emph {et~al.},\ }\href@noop {} {\bibfield  {journal} {\bibinfo
  {journal} {Journal of Physics: Condensed Matter}\ }\textbf {\bibinfo {volume}
  {32}},\ \bibinfo {pages} {193001} (\bibinfo {year} {2020})}\BibitemShut
  {NoStop}%
\bibitem [{\citenamefont {Chat{\'e}}(2020)}]{chate2020dry}%
  \BibitemOpen
  \bibfield  {author} {\bibinfo {author} {\bibfnamefont {H.}~\bibnamefont
  {Chat{\'e}}},\ }\href@noop {} {\bibfield  {journal} {\bibinfo  {journal}
  {Annual Review of Condensed Matter Physics}\ }\textbf {\bibinfo {volume}
  {11}},\ \bibinfo {pages} {189} (\bibinfo {year} {2020})}\BibitemShut
  {NoStop}%
\bibitem [{\citenamefont {Peshkov}\ \emph {et~al.}(2012)\citenamefont
  {Peshkov}, \citenamefont {Aranson}, \citenamefont {Bertin}, \citenamefont
  {Chat{\'e}},\ and\ \citenamefont {Ginelli}}]{peshkov2012nonlinear}%
  \BibitemOpen
  \bibfield  {author} {\bibinfo {author} {\bibfnamefont {A.}~\bibnamefont
  {Peshkov}}, \bibinfo {author} {\bibfnamefont {I.~S.}\ \bibnamefont
  {Aranson}}, \bibinfo {author} {\bibfnamefont {E.}~\bibnamefont {Bertin}},
  \bibinfo {author} {\bibfnamefont {H.}~\bibnamefont {Chat{\'e}}}, \ and\
  \bibinfo {author} {\bibfnamefont {F.}~\bibnamefont {Ginelli}},\ }\href@noop
  {} {\bibfield  {journal} {\bibinfo  {journal} {Physical Review Letters}\
  }\textbf {\bibinfo {volume} {109}},\ \bibinfo {pages} {268701} (\bibinfo
  {year} {2012})}\BibitemShut {NoStop}%
\bibitem [{\citenamefont {Patelli}\ \emph {et~al.}(2019)\citenamefont
  {Patelli}, \citenamefont {Djafer-Cherif}, \citenamefont {Aranson},
  \citenamefont {Bertin},\ and\ \citenamefont
  {Chat{\'e}}}]{patelli2019understanding}%
  \BibitemOpen
  \bibfield  {author} {\bibinfo {author} {\bibfnamefont {A.}~\bibnamefont
  {Patelli}}, \bibinfo {author} {\bibfnamefont {I.}~\bibnamefont
  {Djafer-Cherif}}, \bibinfo {author} {\bibfnamefont {I.~S.}\ \bibnamefont
  {Aranson}}, \bibinfo {author} {\bibfnamefont {E.}~\bibnamefont {Bertin}}, \
  and\ \bibinfo {author} {\bibfnamefont {H.}~\bibnamefont {Chat{\'e}}},\
  }\href@noop {} {\bibfield  {journal} {\bibinfo  {journal} {Physical Review
  Letters}\ }\textbf {\bibinfo {volume} {123}},\ \bibinfo {pages} {258001}
  (\bibinfo {year} {2019})}\BibitemShut {NoStop}%
\bibitem [{\citenamefont {Solon}\ \emph {et~al.}(2015)\citenamefont {Solon},
  \citenamefont {Fily}, \citenamefont {Baskaran}, \citenamefont {Cates},
  \citenamefont {Kafri}, \citenamefont {Kardar},\ and\ \citenamefont
  {Tailleur}}]{solon2015pressure}%
  \BibitemOpen
  \bibfield  {author} {\bibinfo {author} {\bibfnamefont {A.~P.}\ \bibnamefont
  {Solon}}, \bibinfo {author} {\bibfnamefont {Y.}~\bibnamefont {Fily}},
  \bibinfo {author} {\bibfnamefont {A.}~\bibnamefont {Baskaran}}, \bibinfo
  {author} {\bibfnamefont {M.~E.}\ \bibnamefont {Cates}}, \bibinfo {author}
  {\bibfnamefont {Y.}~\bibnamefont {Kafri}}, \bibinfo {author} {\bibfnamefont
  {M.}~\bibnamefont {Kardar}}, \ and\ \bibinfo {author} {\bibfnamefont
  {J.}~\bibnamefont {Tailleur}},\ }\href@noop {} {\bibfield  {journal}
  {\bibinfo  {journal} {Nature Physics}\ }\textbf {\bibinfo {volume} {11}},\
  \bibinfo {pages} {673} (\bibinfo {year} {2015})}\BibitemShut {NoStop}%
\bibitem [{\citenamefont {Alert}\ \emph {et~al.}(2020)\citenamefont {Alert},
  \citenamefont {Joanny},\ and\ \citenamefont
  {Casademunt}}]{alert2020universal}%
  \BibitemOpen
  \bibfield  {author} {\bibinfo {author} {\bibfnamefont {R.}~\bibnamefont
  {Alert}}, \bibinfo {author} {\bibfnamefont {J.-F.}\ \bibnamefont {Joanny}}, \
  and\ \bibinfo {author} {\bibfnamefont {J.}~\bibnamefont {Casademunt}},\
  }\href@noop {} {\bibfield  {journal} {\bibinfo  {journal} {Nature Physics}\
  }\textbf {\bibinfo {volume} {16}},\ \bibinfo {pages} {682} (\bibinfo {year}
  {2020})}\BibitemShut {NoStop}%
\bibitem [{\citenamefont {Peruani}\ and\ \citenamefont
  {Aranson}(2018)}]{peruani2018cold}%
  \BibitemOpen
  \bibfield  {author} {\bibinfo {author} {\bibfnamefont {F.}~\bibnamefont
  {Peruani}}\ and\ \bibinfo {author} {\bibfnamefont {I.~S.}\ \bibnamefont
  {Aranson}},\ }\href@noop {} {\bibfield  {journal} {\bibinfo  {journal}
  {Physical Review Letters}\ }\textbf {\bibinfo {volume} {120}},\ \bibinfo
  {pages} {238101} (\bibinfo {year} {2018})}\BibitemShut {NoStop}%
\bibitem [{\citenamefont {Duan}\ \emph {et~al.}(2021)\citenamefont {Duan},
  \citenamefont {Mahault}, \citenamefont {Ma}, \citenamefont {Shi},\ and\
  \citenamefont {Chat{\'e}}}]{duan2021breakdown}%
  \BibitemOpen
  \bibfield  {author} {\bibinfo {author} {\bibfnamefont {Y.}~\bibnamefont
  {Duan}}, \bibinfo {author} {\bibfnamefont {B.}~\bibnamefont {Mahault}},
  \bibinfo {author} {\bibfnamefont {Y.-q.}\ \bibnamefont {Ma}}, \bibinfo
  {author} {\bibfnamefont {X.-q.}\ \bibnamefont {Shi}}, \ and\ \bibinfo
  {author} {\bibfnamefont {H.}~\bibnamefont {Chat{\'e}}},\ }\href@noop {}
  {\bibfield  {journal} {\bibinfo  {journal} {Physical Review Letters}\
  }\textbf {\bibinfo {volume} {126}},\ \bibinfo {pages} {178001} (\bibinfo
  {year} {2021})}\BibitemShut {NoStop}%
\bibitem [{\citenamefont {Ro}\ \emph {et~al.}(2021)\citenamefont {Ro},
  \citenamefont {Kafri}, \citenamefont {Kardar},\ and\ \citenamefont
  {Tailleur}}]{ro2021disorder}%
  \BibitemOpen
  \bibfield  {author} {\bibinfo {author} {\bibfnamefont {S.}~\bibnamefont
  {Ro}}, \bibinfo {author} {\bibfnamefont {Y.}~\bibnamefont {Kafri}}, \bibinfo
  {author} {\bibfnamefont {M.}~\bibnamefont {Kardar}}, \ and\ \bibinfo {author}
  {\bibfnamefont {J.}~\bibnamefont {Tailleur}},\ }\href@noop {} {\bibfield
  {journal} {\bibinfo  {journal} {Physical Review Letters}\ }\textbf {\bibinfo
  {volume} {126}},\ \bibinfo {pages} {048003} (\bibinfo {year}
  {2021})}\BibitemShut {NoStop}%
\bibitem [{\citenamefont {Olsen}\ \emph {et~al.}(2021)\citenamefont {Olsen},
  \citenamefont {Angheluta},\ and\ \citenamefont
  {Flekk{\o}y}}]{olsen2021active}%
  \BibitemOpen
  \bibfield  {author} {\bibinfo {author} {\bibfnamefont {K.~S.}\ \bibnamefont
  {Olsen}}, \bibinfo {author} {\bibfnamefont {L.}~\bibnamefont {Angheluta}}, \
  and\ \bibinfo {author} {\bibfnamefont {E.~G.}\ \bibnamefont {Flekk{\o}y}},\
  }\href@noop {} {\bibfield  {journal} {\bibinfo  {journal} {Soft Matter}\
  }\textbf {\bibinfo {volume} {17}},\ \bibinfo {pages} {2151} (\bibinfo {year}
  {2021})}\BibitemShut {NoStop}%
\bibitem [{\citenamefont {Waisbord}\ \emph {et~al.}(2021)\citenamefont
  {Waisbord}, \citenamefont {Dehkharghani},\ and\ \citenamefont
  {Guasto}}]{waisbord2021fluidic}%
  \BibitemOpen
  \bibfield  {author} {\bibinfo {author} {\bibfnamefont {N.}~\bibnamefont
  {Waisbord}}, \bibinfo {author} {\bibfnamefont {A.}~\bibnamefont
  {Dehkharghani}}, \ and\ \bibinfo {author} {\bibfnamefont {J.~S.}\
  \bibnamefont {Guasto}},\ }\href@noop {} {\bibfield  {journal} {\bibinfo
  {journal} {Nature Communications}\ }\textbf {\bibinfo {volume} {12}},\
  \bibinfo {pages} {1} (\bibinfo {year} {2021})}\BibitemShut {NoStop}%
\bibitem [{\citenamefont {Bychkov}\ and\ \citenamefont
  {Dykhne}(1971)}]{bychkov1971theory}%
  \BibitemOpen
  \bibfield  {author} {\bibinfo {author} {\bibfnamefont {Y.~A.}\ \bibnamefont
  {Bychkov}}\ and\ \bibinfo {author} {\bibfnamefont {A.~M.}\ \bibnamefont
  {Dykhne}},\ }\href@noop {} {\bibfield  {journal} {\bibinfo  {journal}
  {Theoretical and Mathematical Physics}\ }\textbf {\bibinfo {volume} {6}},\
  \bibinfo {pages} {307} (\bibinfo {year} {1971})}\BibitemShut {NoStop}%
\bibitem [{\citenamefont {Dauchot}\ and\ \citenamefont
  {D{\'e}mery}(2019)}]{dauchot2019dynamics}%
  \BibitemOpen
  \bibfield  {author} {\bibinfo {author} {\bibfnamefont {O.}~\bibnamefont
  {Dauchot}}\ and\ \bibinfo {author} {\bibfnamefont {V.}~\bibnamefont
  {D{\'e}mery}},\ }\href@noop {} {\bibfield  {journal} {\bibinfo  {journal}
  {Physical Review Letters}\ }\textbf {\bibinfo {volume} {122}},\ \bibinfo
  {pages} {068002} (\bibinfo {year} {2019})}\BibitemShut {NoStop}%
\bibitem [{\citenamefont {Schmidt}\ \emph {et~al.}(2021)\citenamefont
  {Schmidt}, \citenamefont {{\v{S}}{\'\i}pov{\'a}-Jungov{\'a}}, \citenamefont
  {K{\"a}ll}, \citenamefont {W{\"u}rger},\ and\ \citenamefont
  {Volpe}}]{schmidt2021non}%
  \BibitemOpen
  \bibfield  {author} {\bibinfo {author} {\bibfnamefont {F.}~\bibnamefont
  {Schmidt}}, \bibinfo {author} {\bibfnamefont {H.}~\bibnamefont
  {{\v{S}}{\'\i}pov{\'a}-Jungov{\'a}}}, \bibinfo {author} {\bibfnamefont
  {M.}~\bibnamefont {K{\"a}ll}}, \bibinfo {author} {\bibfnamefont
  {A.}~\bibnamefont {W{\"u}rger}}, \ and\ \bibinfo {author} {\bibfnamefont
  {G.}~\bibnamefont {Volpe}},\ }\href@noop {} {\bibfield  {journal} {\bibinfo
  {journal} {Nature Communications}\ }\textbf {\bibinfo {volume} {12}},\
  \bibinfo {pages} {1} (\bibinfo {year} {2021})}\BibitemShut {NoStop}%
\bibitem [{\citenamefont {Takatori}\ \emph {et~al.}(2016)\citenamefont
  {Takatori}, \citenamefont {De~Dier}, \citenamefont {Vermant},\ and\
  \citenamefont {Brady}}]{takatori2016acoustic}%
  \BibitemOpen
  \bibfield  {author} {\bibinfo {author} {\bibfnamefont {S.~C.}\ \bibnamefont
  {Takatori}}, \bibinfo {author} {\bibfnamefont {R.}~\bibnamefont {De~Dier}},
  \bibinfo {author} {\bibfnamefont {J.}~\bibnamefont {Vermant}}, \ and\
  \bibinfo {author} {\bibfnamefont {J.~F.}\ \bibnamefont {Brady}},\ }\href@noop
  {} {\bibfield  {journal} {\bibinfo  {journal} {Nature Communications}\
  }\textbf {\bibinfo {volume} {7}},\ \bibinfo {pages} {1} (\bibinfo {year}
  {2016})}\BibitemShut {NoStop}%
\bibitem [{\citenamefont {Pototsky}\ and\ \citenamefont
  {Stark}(2012)}]{pototsky2012active}%
  \BibitemOpen
  \bibfield  {author} {\bibinfo {author} {\bibfnamefont {A.}~\bibnamefont
  {Pototsky}}\ and\ \bibinfo {author} {\bibfnamefont {H.}~\bibnamefont
  {Stark}},\ }\href@noop {} {\bibfield  {journal} {\bibinfo  {journal} {EPL}\
  }\textbf {\bibinfo {volume} {98}},\ \bibinfo {pages} {50004} (\bibinfo {year}
  {2012})}\BibitemShut {NoStop}%
\bibitem [{\citenamefont {Hennes}\ \emph {et~al.}(2014)\citenamefont {Hennes},
  \citenamefont {Wolff},\ and\ \citenamefont {Stark}}]{hennes2014self}%
  \BibitemOpen
  \bibfield  {author} {\bibinfo {author} {\bibfnamefont {M.}~\bibnamefont
  {Hennes}}, \bibinfo {author} {\bibfnamefont {K.}~\bibnamefont {Wolff}}, \
  and\ \bibinfo {author} {\bibfnamefont {H.}~\bibnamefont {Stark}},\
  }\href@noop {} {\bibfield  {journal} {\bibinfo  {journal} {Physical Review
  Letters}\ }\textbf {\bibinfo {volume} {112}},\ \bibinfo {pages} {238104}
  (\bibinfo {year} {2014})}\BibitemShut {NoStop}%
\bibitem [{\citenamefont {Wexler}\ \emph {et~al.}(2020)\citenamefont {Wexler},
  \citenamefont {Gov}, \citenamefont {Rasmussen},\ and\ \citenamefont
  {Bel}}]{wexler2020dynamics}%
  \BibitemOpen
  \bibfield  {author} {\bibinfo {author} {\bibfnamefont {D.}~\bibnamefont
  {Wexler}}, \bibinfo {author} {\bibfnamefont {N.}~\bibnamefont {Gov}},
  \bibinfo {author} {\bibfnamefont {K.~{\O}.}\ \bibnamefont {Rasmussen}}, \
  and\ \bibinfo {author} {\bibfnamefont {G.}~\bibnamefont {Bel}},\ }\href@noop
  {} {\bibfield  {journal} {\bibinfo  {journal} {Physical Review Research}\
  }\textbf {\bibinfo {volume} {2}},\ \bibinfo {pages} {013003} (\bibinfo {year}
  {2020})}\BibitemShut {NoStop}%
\bibitem [{\citenamefont {Das}\ \emph {et~al.}(2018)\citenamefont {Das},
  \citenamefont {Gompper},\ and\ \citenamefont {Winkler}}]{das2018confined}%
  \BibitemOpen
  \bibfield  {author} {\bibinfo {author} {\bibfnamefont {S.}~\bibnamefont
  {Das}}, \bibinfo {author} {\bibfnamefont {G.}~\bibnamefont {Gompper}}, \ and\
  \bibinfo {author} {\bibfnamefont {R.~G.}\ \bibnamefont {Winkler}},\ }\href
  {\doibase 10.1088/1367-2630/aa9d4b} {\bibfield  {journal} {\bibinfo
  {journal} {New Journal of Physics}\ }\textbf {\bibinfo {volume} {20}},\
  \bibinfo {pages} {015001} (\bibinfo {year} {2018})}\BibitemShut {NoStop}%
\bibitem [{\citenamefont {Jahanshahi}\ \emph {et~al.}(2017)\citenamefont
  {Jahanshahi}, \citenamefont {L{\"o}wen},\ and\ \citenamefont
  {Ten~Hagen}}]{jahanshahi2017brownian}%
  \BibitemOpen
  \bibfield  {author} {\bibinfo {author} {\bibfnamefont {S.}~\bibnamefont
  {Jahanshahi}}, \bibinfo {author} {\bibfnamefont {H.}~\bibnamefont
  {L{\"o}wen}}, \ and\ \bibinfo {author} {\bibfnamefont {B.}~\bibnamefont
  {Ten~Hagen}},\ }\href@noop {} {\bibfield  {journal} {\bibinfo  {journal}
  {Physical Review E}\ }\textbf {\bibinfo {volume} {95}},\ \bibinfo {pages}
  {022606} (\bibinfo {year} {2017})}\BibitemShut {NoStop}%
\bibitem [{\citenamefont {Liebchen}\ \emph {et~al.}(2016)\citenamefont
  {Liebchen}, \citenamefont {Cates},\ and\ \citenamefont
  {Marenduzzo}}]{liebchen2016pattern}%
  \BibitemOpen
  \bibfield  {author} {\bibinfo {author} {\bibfnamefont {B.}~\bibnamefont
  {Liebchen}}, \bibinfo {author} {\bibfnamefont {M.~E.}\ \bibnamefont {Cates}},
  \ and\ \bibinfo {author} {\bibfnamefont {D.}~\bibnamefont {Marenduzzo}},\
  }\href@noop {} {\bibfield  {journal} {\bibinfo  {journal} {Soft Matter}\
  }\textbf {\bibinfo {volume} {12}},\ \bibinfo {pages} {7259} (\bibinfo {year}
  {2016})}\BibitemShut {NoStop}%
\bibitem [{\citenamefont {Liebchen}\ and\ \citenamefont
  {L\"owen}(2018)}]{liebchen2018synthetic}%
  \BibitemOpen
  \bibfield  {author} {\bibinfo {author} {\bibfnamefont {B.}~\bibnamefont
  {Liebchen}}\ and\ \bibinfo {author} {\bibfnamefont {H.}~\bibnamefont
  {L\"owen}},\ }\href@noop {} {\bibfield  {journal} {\bibinfo  {journal}
  {Accounts of Chemical Research}\ }\textbf {\bibinfo {volume} {51}},\ \bibinfo
  {pages} {2982} (\bibinfo {year} {2018})}\BibitemShut {NoStop}%
\bibitem [{\citenamefont {Stark}(2018)}]{stark2018artificial}%
  \BibitemOpen
  \bibfield  {author} {\bibinfo {author} {\bibfnamefont {H.}~\bibnamefont
  {Stark}},\ }\href@noop {} {\bibfield  {journal} {\bibinfo  {journal}
  {Accounts of Chemical Research}\ }\textbf {\bibinfo {volume} {51}},\ \bibinfo
  {pages} {2681} (\bibinfo {year} {2018})}\BibitemShut {NoStop}%
\bibitem [{\citenamefont {Kravtsov}\ and\ \citenamefont
  {Orlov}(1990)}]{Kravtsov-Orlov-90}%
  \BibitemOpen
  \bibfield  {author} {\bibinfo {author} {\bibfnamefont {Y.~A.}\ \bibnamefont
  {Kravtsov}}\ and\ \bibinfo {author} {\bibfnamefont {Y.~I.}\ \bibnamefont
  {Orlov}},\ }\href@noop {} {\emph {\bibinfo {title} {Geometrical Optics of
  Inhomogeneous Media}}}\ (\bibinfo  {publisher} {Springer},\ \bibinfo
  {address} {Berlin, Heidelberg},\ \bibinfo {year} {1990})\BibitemShut
  {NoStop}%
\bibitem [{sup()}]{suppl}%
  \BibitemOpen
  \href@noop {} {\bibinfo  {journal} {See Supplemental Material at
  $http://link.aps.org/supplemental/$ for computational videos}\ }\BibitemShut
  {NoStop}%
\bibitem [{\citenamefont {Vicsek}\ \emph {et~al.}(1995)\citenamefont {Vicsek},
  \citenamefont {Czir\'ok}, \citenamefont {Ben-Jacob}, \citenamefont {Cohen},\
  and\ \citenamefont {Shochet}}]{Vicsek_etal-95}%
  \BibitemOpen
\bibfield  {journal} {  }\bibfield  {author} {\bibinfo {author} {\bibfnamefont
  {T.}~\bibnamefont {Vicsek}}, \bibinfo {author} {\bibfnamefont
  {A.}~\bibnamefont {Czir\'ok}}, \bibinfo {author} {\bibfnamefont
  {E.}~\bibnamefont {Ben-Jacob}}, \bibinfo {author} {\bibfnamefont
  {I.}~\bibnamefont {Cohen}}, \ and\ \bibinfo {author} {\bibfnamefont
  {O.}~\bibnamefont {Shochet}},\ }\href@noop {} {\bibfield  {journal} {\bibinfo
   {journal} {Phys. Rev. Lett.}\ }\textbf {\bibinfo {volume} {75}},\ \bibinfo
  {pages} {1226} (\bibinfo {year} {1995})}\BibitemShut {NoStop}%
\bibitem [{\citenamefont {Kruk}\ \emph {et~al.}(2018)\citenamefont {Kruk},
  \citenamefont {Maistrenko},\ and\ \citenamefont {Koeppl}}]{Kruk_etal-18}%
  \BibitemOpen
  \bibfield  {author} {\bibinfo {author} {\bibfnamefont {N.}~\bibnamefont
  {Kruk}}, \bibinfo {author} {\bibfnamefont {Y.}~\bibnamefont {Maistrenko}}, \
  and\ \bibinfo {author} {\bibfnamefont {H.}~\bibnamefont {Koeppl}},\
  }\href@noop {} {\bibfield  {journal} {\bibinfo  {journal} {Phys. Rev. E}\
  }\textbf {\bibinfo {volume} {98}},\ \bibinfo {pages} {032219} (\bibinfo
  {year} {2018})}\BibitemShut {NoStop}%
\bibitem [{\citenamefont {Pecora}\ and\ \citenamefont
  {Carroll}(1998)}]{pecora1998master}%
  \BibitemOpen
  \bibfield  {author} {\bibinfo {author} {\bibfnamefont {L.~M.}\ \bibnamefont
  {Pecora}}\ and\ \bibinfo {author} {\bibfnamefont {T.~L.}\ \bibnamefont
  {Carroll}},\ }\href@noop {} {\bibfield  {journal} {\bibinfo  {journal}
  {Physical Review Letters}\ }\textbf {\bibinfo {volume} {80}},\ \bibinfo
  {pages} {2109} (\bibinfo {year} {1998})}\BibitemShut {NoStop}%
\bibitem [{\citenamefont {Lai}\ and\ \citenamefont
  {T{\'e}l}(2011)}]{Lai-Tel-11}%
  \BibitemOpen
  \bibfield  {author} {\bibinfo {author} {\bibfnamefont {Y.-C.}\ \bibnamefont
  {Lai}}\ and\ \bibinfo {author} {\bibfnamefont {T.}~\bibnamefont {T{\'e}l}},\
  }\href@noop {} {\emph {\bibinfo {title} {Transient Chaos}}}\ (\bibinfo
  {publisher} {Springer},\ \bibinfo {address} {New York},\ \bibinfo {year}
  {2011})\BibitemShut {NoStop}%
\bibitem [{\citenamefont {Pikovsky}(2021)}]{Pikovsky-21a}%
  \BibitemOpen
  \bibfield  {author} {\bibinfo {author} {\bibfnamefont {A.}~\bibnamefont
  {Pikovsky}},\ }\href@noop {} {\bibfield  {journal} {\bibinfo  {journal} {J.
  Phys. Complexity}\ }\textbf {\bibinfo {volume} {2}},\ \bibinfo {pages}
  {025009} (\bibinfo {year} {2021})}\BibitemShut {NoStop}%
\bibitem [{\citenamefont {Bak}\ \emph {et~al.}(1988)\citenamefont {Bak},
  \citenamefont {Tang},\ and\ \citenamefont {Wiesenfeld}}]{bak1988self}%
  \BibitemOpen
  \bibfield  {author} {\bibinfo {author} {\bibfnamefont {P.}~\bibnamefont
  {Bak}}, \bibinfo {author} {\bibfnamefont {C.}~\bibnamefont {Tang}}, \ and\
  \bibinfo {author} {\bibfnamefont {K.}~\bibnamefont {Wiesenfeld}},\
  }\href@noop {} {\bibfield  {journal} {\bibinfo  {journal} {Physical Review
  A}\ }\textbf {\bibinfo {volume} {38}},\ \bibinfo {pages} {364} (\bibinfo
  {year} {1988})}\BibitemShut {NoStop}%
\end{thebibliography}

\end{document}